\documentclass[journal,10pt]{IEEEtran}
\usepackage{cite}
\usepackage{bbding}
\usepackage{makecell}
\usepackage{graphicx}
\usepackage{float}
\usepackage{amsthm}
\usepackage[colorlinks=black,linkcolor=black,anchorcolor=black,citecolor=black]{hyperref}
\usepackage{amsmath,amssymb,amsfonts,bm,subfigure,lipsum,amsthm}
\usepackage{stfloats}
\usepackage{algorithm}
\usepackage{algorithmic}
\usepackage{mathrsfs}
\usepackage{graphicx}
\usepackage[table]{xcolor}
\usepackage{threeparttable}
\usepackage{textcomp}
\usepackage{xcolor}
\usepackage{color}
\usepackage{booktabs}
\usepackage{flushend}
\usepackage{amsmath}
\usepackage{bbm}
\usepackage{multicol}
\usepackage{multirow}
\usepackage[english]{babel}
\usepackage{upgreek}
\usepackage{float}
\usepackage{colortbl}
\usepackage[table]{xcolor}
\usepackage{cuted}
\usepackage{epstopdf}

\def\BibTeX{{\rm B\kern-.05em{\sc i\kern-.025em b}\kern-.08em
    T\kern-.1667em\lower.7ex\hbox{E}\kern-.125emX}}
\usepackage{graphicx,graphics,color,epsfig,subfigure,graphpap,rotate}
\usepackage{times, verbatim, subfigure, epsfig, graphicx, latexsym}
\usepackage{url}
\usepackage{subfigure}

\definecolor{hellmagenta}{rgb}{1,0.75,0.9}

\definecolor{hellcyan}{rgb}{0.75,1,0.9}

\definecolor{hellgelb}{rgb}{1,1,0.8}

\definecolor{colKeys}{rgb}{0,0,1}

\definecolor{colIdentifier}{rgb}{0,0,0}

\definecolor{colComments}{rgb}{1,0,0}

\definecolor{colString}{rgb}{0,0.5,0}

\definecolor{darkyellow}{rgb}{1,0.9,0}

\begin{document}
\title{ Towards Secure  Semantic Transmission In the Era of GenAI: A Diffusion-based Framework}

\author{Boxiang~He,~Zihan~Chen,~Junshan~Luo,~Chuanhong~Liu,~Shilian~Wang,\\~Fanggang~Wang,~\IEEEmembership{Senior~Member,~IEEE},~and~Tony~Q.S.~Quek,~\IEEEmembership{Fellow,~IEEE}

\thanks{ Boxiang  He, Junshan Luo, and Shilian Wang are with the
College of Electronic Science and Technology, National University of Defense Technology, Changsha 410073, China (e-mail: boxianghe1@bjtu.edu.cn; ljsnudt@foxmail.com; wangsl@nudt.edu.cn).

Zihan Chen and Tony Q.S. Quek are with the  Information Systems Technology and
Design, Singapore University of Technology and Design, Singapore
(e-mail: zihan$\_$chen@mymail.sutd.edu.sg; tonyquek@sutd.edu.sg).

Chuanhong Liu is with the Beijing Key Laboratory
of Network System Architecture and Convergence, School of Information
and Communication Engineering, Beijing University of Posts and
Telecommunications, Beijing 100876, China (e-mail: 2016$\_$liuchuanhong@bupt.edu.cn).

Fanggang Wang is with the School of Electronic and Information Engineering, Beijing Jiaotong University, Beijing 100044, China (e-mail:  wangfg@bjtu.edu.cn).

}

}

\maketitle

\begin{abstract}
Semantic communication, due to its focus on the transmitting meaning rather than the raw bit data, poses unique security challenges compared to the traditional communication systems. In particular, semantic communication systems are vulnerable to the malicious attacks that focus on the semantic layer, with the intention of understanding or distorting the intended meaning of the transmitted privacy data. Diffusion models, a class of generative artificial intelligence (GenAI), are well-suited for ensuring data security to attack. Through iteratively adding and then removing noise, diffusion models can generate meaningful information despite the presence of the  unknown noise. This article proposes a diffusion-based framework to enhance the security of semantic transmission  for the attacks including  eavesdropping and jamming.  Specifically, the proposed framework incorporates both the artificial noise  and natural channel noise into the forward process of the diffusion models during the semantic transmission, with the reverse process used to remove noise at the legitimate receiver. In the  eavesdropping scenarios, the artificial  noise is the friendly noise designed to prevent semantic eavesdropping. In the  jamming scenarios, the artificial noise is  the malicious jamming generated by the jammer, which disrupts the semantic transmission. The case studies show that the proposed diffusion-based framework is promising in securing the semantic transmission. We also consolidate several broad research directions associated with the proposed  framework.

\end{abstract}


\section{Introduction}

\IEEEPARstart{T}{he} sixth generation (6G) networks are expected to enable the ultra-high-speed data transmission, the extensive network coverage, and the ultra-low latency, which can provide the strong support for the applications such as the autonomous driving, the smart cities, and the remote healthcare\cite{Saad_network}. As the number of connected devices increases and the application scenarios become more complex, the traditional bit-based communication technologies face the significant challenges in terms of efficiency and latency. In this context, semantic communication emerges as a promising solution. By transmitting the information meaning  rather than the  bit streams, semantic communication can significantly enhance  efficiency and reduce latency, thus offering a more intelligent and efficient solution for 6G networks\cite{luo2022Se,yang2023se,qin2022se}.

However, semantic communication presents unique challenges in ensuring security compared to the traditional bit communication\cite{yang_network,shen_network}. Specifically, the traditional communication systems focus on protecting the  security of raw data transmitted as bit streams, while semantic communication prioritizes the secure transmission of meaningful information, often relying on context, intent, and content recognition. This shift introduces new vulnerabilities, as attackers may target the underlying semantic layers, aiming to understand or distort the intended meaning of the transmitted privacy information. In this context, traditional security mechanisms such as encryption and hashing are insufficient, necessitating the development of novel approaches tailored to the unique characteristics of semantic transmission.

\begin{figure*}[tbp]
  \centering
  \includegraphics[width=0.5\linewidth]{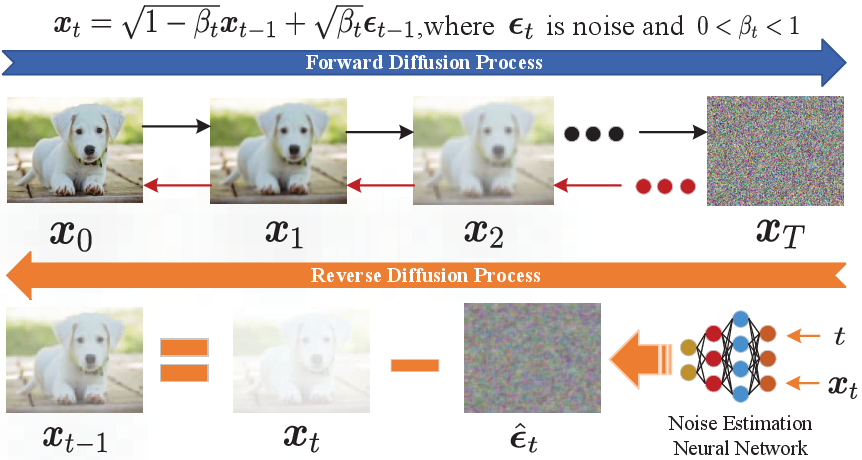}\\
  \caption{ Overview of the diffusion models.}  \label{diffusion}
\end{figure*}

Among the emerging technologies for enhancing the security of semantic communication, diffusion models have shown considerable promise. In the era of generative artificial intelligence (GenAI), diffusion models represent a new class of generative models that can generate high-quality data by learning to iteratively add noise to the data and then reverse the process to generate the original data\cite{ho2020denoising}.  The reverse process makes diffusion models particularly suitable for ensuring data robustness to noise, which is essential in the security of semantic transmission. The benefits of  diffusion models in securing semantic transmission are  discussed below.

\begin{itemize}
  \item  {\textbf{Robustness:}}
One of the primary advantages of diffusion models is the robustness to noise. In semantic transmission,  both artificial noise and natural  noise can corrupt the transmitted data, making it difficult to preserve the  security of semantic information. In contrast, diffusion models are designed to progressively denoising data, allowing the system to generate meaningful content even in the presence of noise. This property is particularly useful for  the security of semantic transmission.

  \item  {\textbf{Stability:}} Another key strength of diffusion models is the stability compared to other generative models such as generative adversarial networks (GANs)\cite{zou_gen_magazine}. GANs  often suffer from issues like mode collapse or training instability, which can affect the quality  of generated data. Diffusion models, on the other hand, are less prone to such issues, offering a more stable  framework for generating high-quality  representations of data \cite{letafati2023diffusionmodelswirelesscommunications}. This stability is essential for maintaining the security of semantic information, where any deviation from the intended message could lead to severe consequences, especially in critical applications such as autonomous systems or remote medical operations.
\end{itemize}

Building on these strengths, this paper proposes a diffusion-based framework for ensuring the security of semantic transmission. By leveraging the inherent properties of diffusion models, we aim to provide a comprehensive solution for securing the transmission of semantic data in next-generation communication systems.  \emph{The proposed framework focuses on exploring how diffusion models can be employed to address the attack issues in semantic transmission.} In particular, Section \ref{SectionII} introduces the preliminaries of diffusion models. Section \ref{SectionIII} identifies the potential attacks of semantic transmission from the  perspective of eavesdropping and jamming. In Section \ref{SectionIV}, the eavesdropping-aware semantic communication framework via diffusion models is designed. Section \ref{SectionV} proposes the jamming-aware semantic communication framework via diffusion models. Section \ref{SectionVI} discusses the potential challenges and future directions for the proposed framework.

\section{Preliminaries of  Diffusion Models} \label{SectionII}

In this section, we introduce the preliminaries of diffusion models. Diffusion models are a class of generative models that generate data by simulating a diffusion process. As shown in Fig. \ref{diffusion}, the diffusion models involve two key phases, i.e., forward diffusion process and reverse diffusion process.

The forward diffusion process is essentially a Markov chain where each time-step in the chain adds noise to the data, typically in the form of Gaussian noise. Starting from the original data, a small amount of noise is added at each time step until the data is eventually transformed into a noisy representation, which is often indistinguishable from random noise. The forward process gradually makes the data less identifiable, effectively corrupting data over time. The amount of noise added at each time-step is determined by a variance schedule $\beta_t$, which specifies how much noise should be added during each step. The schedule can vary, but typically follows a gradual increase in noise as the process progresses. The forward diffusion process can be seen as a sequence of steps that slowly degrade the data, making it harder to reconstruct without knowing the exact sequence of noise additions. One of the advantages of this approach is that it allows for a well-defined, predictable process for generating noise-corrupted data, which is essential for training the reverse process.

The reverse diffusion process is where the actual generative model is learned. The goal of reverse diffusion process is to reverse the forward diffusion process and recover the original data from its noisy version. The reverse process works by gradually removing the noise added in the forward process, step-by-step, and attempting to reconstruct the original data. This is achieved by learning a noise estimation neural network  that approximates the noise added at each corresponding time-step of the forward process. The noise estimation neural network is trained to predict the noise added at each time-step and use the estimated noise to  recover the original data. During training, the noise estimation neural network is optimized to minimize the difference between the predicted noise and the actual noise that is added in the forward process.

The diffusion models represent a significant departure from traditional GenAI, which mainly includes GANs and Autoencoders (AEs). Unlike GANs, which aim to directly learn the data distribution, or AEs, which focus on learning a latent space embedding, diffusion models work by progressively adding noise to data samples. The models then attempt to recover the original data by denoising the perturbed samples. Thus,  diffusion models have two main advantages in robustness and stability. For robustness, diffusion models are capable of generating high-fidelity data in the presence of  noise. Since the process involves multiple steps of fine-grained denoising, the model can generate data with intricate details and high quality. This has made diffusion models the method of choice for tasks such as image generation and enhancement. For stability,  compared to other generative models, such as GANs,  training can be unstable due to the adversarial nature of the setup, where the generator and discriminator must compete. In contrast, diffusion models avoid such instabilities by treating the problem as a denoising task, which is easier to optimize.

\begin{figure*}[tbp]
  \centering
  \includegraphics[width=5.5in]{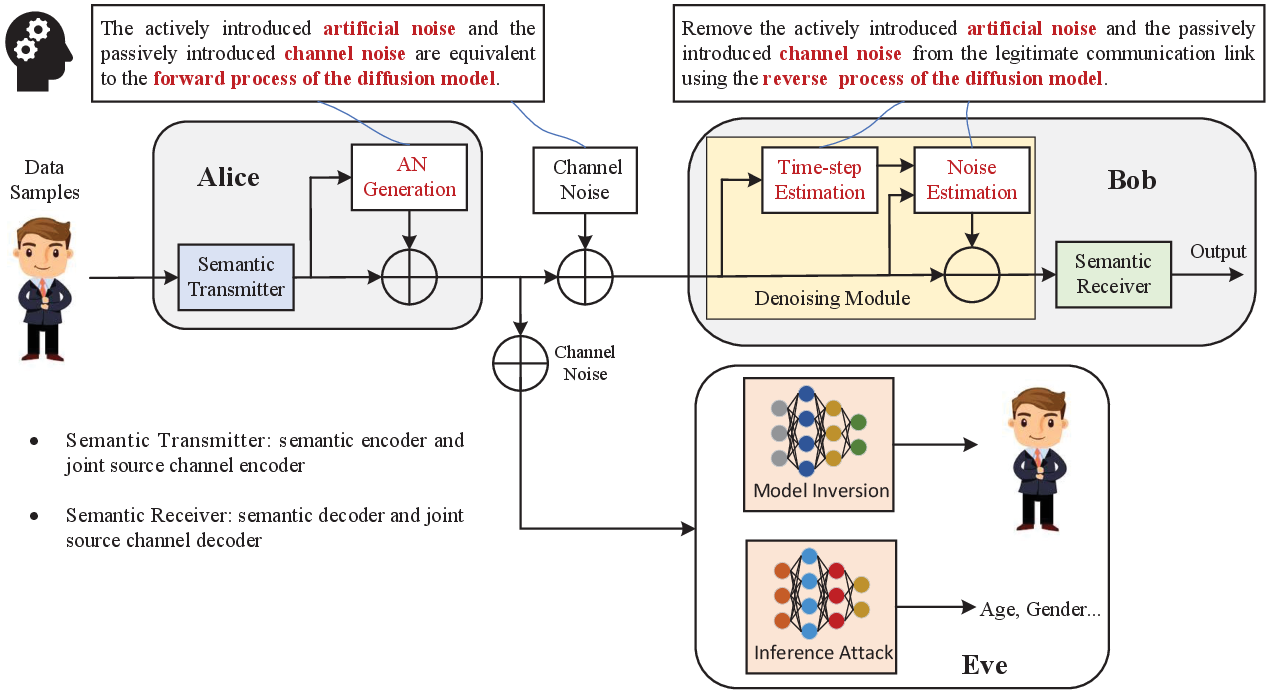}\\
  \caption{  Eavesdropping-aware semantic
communication framework via diffusion models.}  \label{SC_eavesdropping}
\end{figure*}

\section{Attack Risks of Semantic Transmission} \label{SectionIII}

In this section, we present a balanced discussion on attack risks in the stage of semantic transmission, which encompasses two categories, i.e., eavesdropping and jamming. It is worth noting that various types of attacks during the semantic transmission phase can be classified into the two categories mentioned above.

\subsection{Eavesdropping Attack for Semantic Communication}
Eavesdropping  refers to the interception and interpretation of the transmitted message by an unauthorized party, aiming to extract privacy information from the content of the legitimate communication.  Generally, there are two categories of eavesdropping in semantic transmission: Model inversion attack\cite{chen_the_model} and inference attack\cite{liu2023sem,he2024secure}.

\subsubsection{Model Inversion Attack}
The model inversion attack is a type of attack on deep learning models, where the eavesdropper attempts to infer the sensitive information of  data by reverse-engineering the  model. In semantic communication system,  the eavesdropper typically tries to reconstruct the original data by using the model inversion attack, in which the eavesdropping process is  as follows: Alice and Bob deploy the trained semantic transmitter and semantic receiver respectively. At the same time, the trained models are released for public usage. Thus, when eavesdropper Eve receives the signal transmitted by Alice, Eve can acquire the recovered data by
inverting the public models.

%

%
%
%
%

\subsubsection{Inference Attack} Unlike model inversion attacks, which focus on reconstructing the original data, the inference attack aims to deduce the sensitive information by analyzing the outputs or behavior of the deep learning models.  In semantic communication system, once Eve receives the signal sent by Alice to Bob, Eve determines the private information in the transmitted signal of Alice by using the inference attack.
Common inference attack can be categorized into two main types: Membership inference attack and attribute inference attack. For  the membership inference attack, the eavesdropper attempts to determine whether a specific individual or item is a member of a certain dataset. For example, Eve may try to figure out if someone is a patient of a particular hospital. For the attribute inference attack, the eavesdropper attempts to infer private attributes  like age, gender, or health status.

%
%

%
%

\subsection{Jamming Attack for Semantic Communication} \label{subsection_JSC}

Jamming  refers to intentional interference or disruption of the communication channel to degrade or prevent the receiver from correctly interpreting the transmitted message.  In the context of semantic communication, jamming includes two types: High-power jamming \cite{pri_jamming} and low-power adversarial attack\cite{madry2017towards}.

\subsubsection{High-power Jamming} High-power jamming is a form of deliberate interference, in which a jammer transmits a signal at a much higher power level than the legitimate communication system. The goal is to overpower the target signal and render the communication unreliable or entirely inoperative. Common high-power jamming includes continuous wave jamming, random noise jamming, pulse jamming, reactive jamming, and frequency sweeping jamming. In semantic communication system, once the jammer transmits the high-power jamming signal, it will be difficult for the semantic receiver to understand the meaning of the transmitted signal due to the high jamming-to-signal ratio (JSR).
\begin{figure*}[tbp]
  \centering
  \includegraphics[width=5.6in]{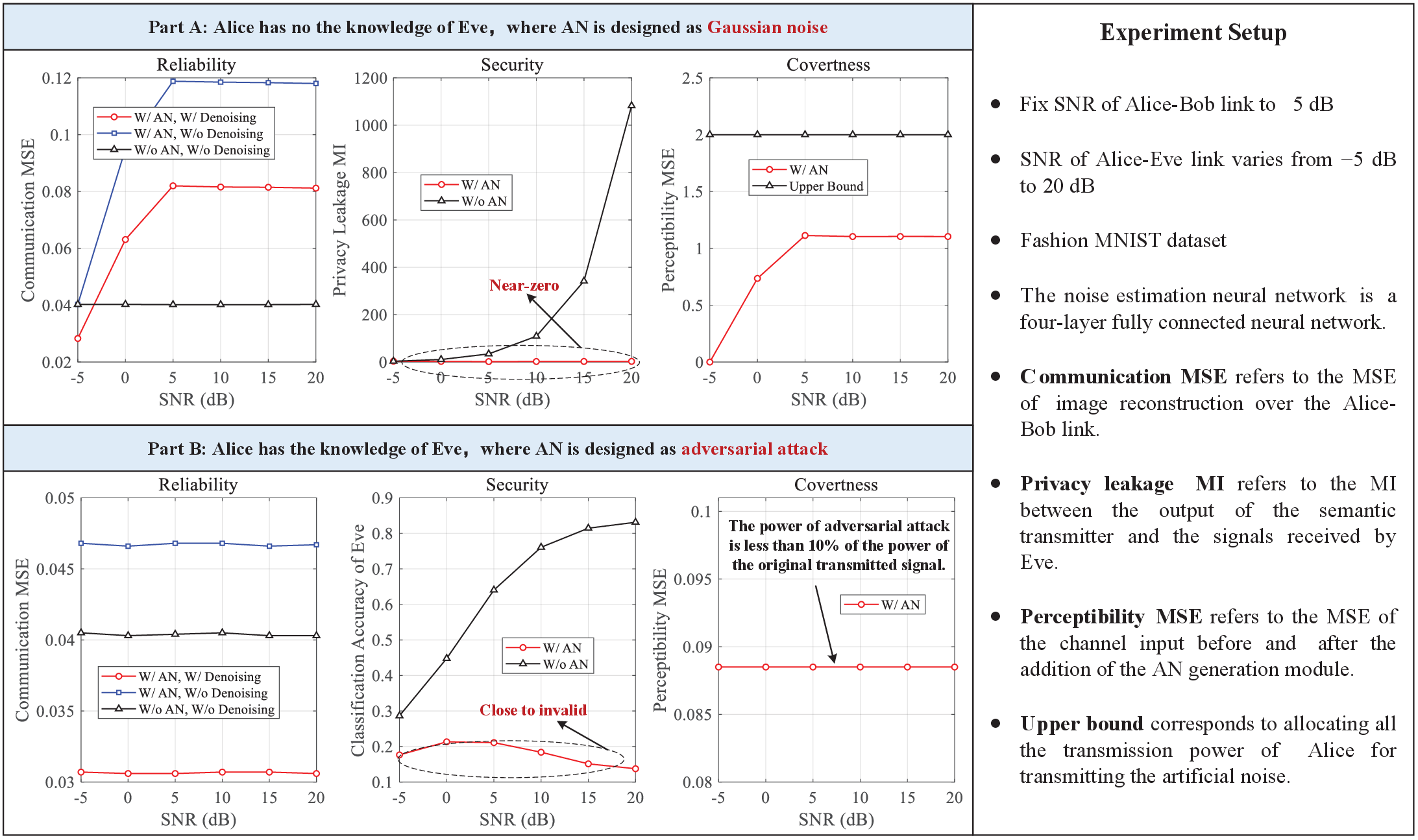}\\
  \caption{  Performance evaluation of the proposed eavesdropping-aware semantic communication framework via diffusion models.}  \label{fashion_eavesdropping}
\end{figure*}

\subsubsection{Low-power Adversarial Attack}  Adversarial attack exploits the vulnerabilities of deep learning models, where  the  imperceptible changes to the input data can cause the model to misinterpret or incorrectly classify the data. Common adversarial attack generation methods include fast gradient sign method, basic iterative method, and projected gradient descent. In semantic communication, the jammer launches the low-power adversarial attacks that are superimposed on Alice's transmitted signals. These subtle adversarial attacks can confuse the semantic receiver at Bob, leading to incorrect interpretation of the message.

%


\section{Eavesdropping-aware Semantic Communication Framework Via Diffusion Models} \label{SectionIV}

To address the eavesdropping problem of  semantic communication, as shown in Fig. \ref{SC_eavesdropping}, we propose a novel eavesdropping-aware semantic communication framework mainly consists of two modules: The AN generation module and the denoising module. The AN generation module generates AN to prevent Eve from illegally eavesdropping, while the denoising module removes the unwanted noise at  Bob that reduces the reliability of legitimate communications.

\subsection{AN Generation Module}

The primary goal of the AN generation module is to inject AN into the output of semantic transmitter to confuse  eavesdropper, while maintaining the high-quality semantic communication. To this end, two  steps need to be  designed including power allocation and  noise design.

\begin{figure*}[tbp]
  \centering
  \includegraphics[width=6.5in]{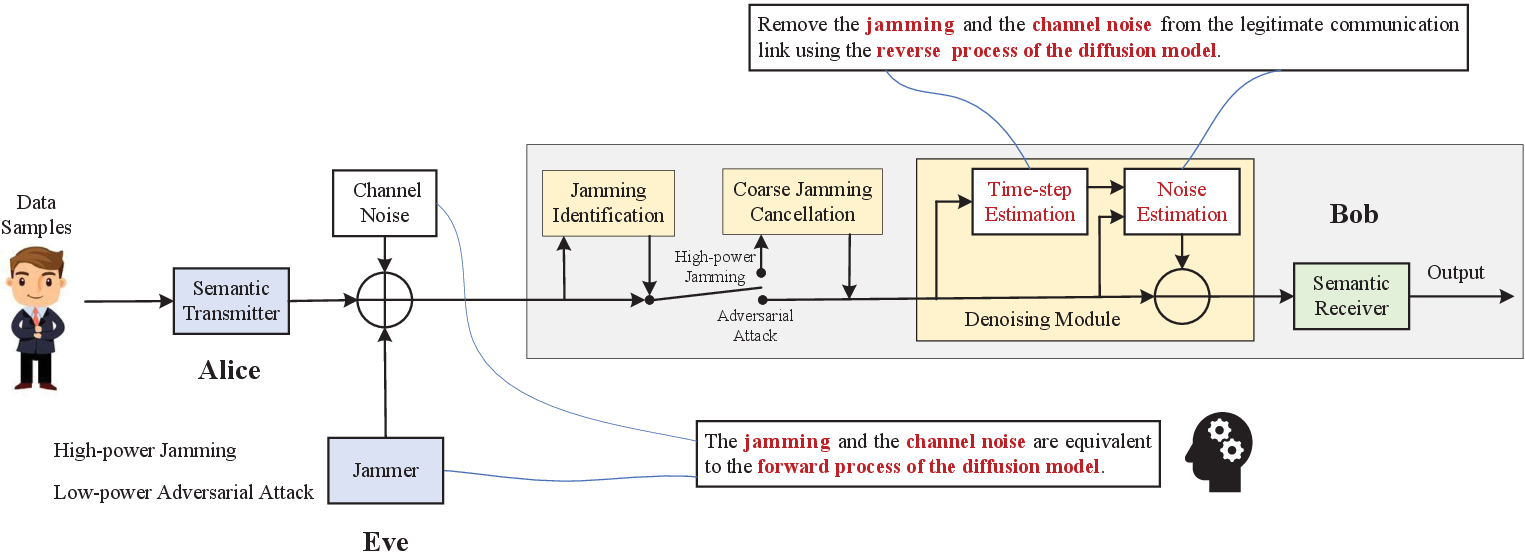}\\
  \caption{  Jamming-aware semantic
communication framework via diffusion models.}  \label{SC_jamming}
\end{figure*}

\subsubsection{Power Allocation}
The most crucial aspect of the AN generation module is power allocation between the output of semantic transmitter and the artificial noise. In semantic communication system, the power allocated to the noise must be carefully chosen to ensure the following:

\begin{itemize}
\item {\textbf{Security:}} {The noise power is strong enough to significantly degrade Eve's ability to   extract the privacy information. The greater the noise power, the less information leaked to Eve. }
\item {\textbf{Reliability:}} {The power allocated to the legitimate signal is sufficient for Bob to reliably  interpret the transmitted message of Alice. The smaller the noise power, the better Bob can perform semantic decoding.}
\item {\textbf{Covertness:}} {The inserted noise should not cause over-distortion of the channel input. In other words, the output of the semantic transmitter should remain as unchanged as possible before and after the noise is added, which can reduce the suspicion of the eavesdropper. }
\end{itemize}

Thus, researchers can first design reasonable metrics to characterize the security, the reliability,  and the covertness, and then solve the optimal power allocation  by maximizing or minimizing the weighted sum of these metrics. The typical solution method is the reinforcement learning.

\subsubsection{Artificial Noise Design}  In addition to power allocation, AN design also affects the trade-offs between security, reliability, and covertness in semantic communication systems. We discuss the following two cases based on whether Alice has the knowledge of Eve.\footnote{In the semantic communication system, Eve's knowledge refers to the deep learning model, parameters, and specific eavesdropping tasks used by Eve.}
{When Alice has no the knowledge of Eve, \emph{\textbf{AN can be designed as Gaussian noise}}. This means that Alice only needs to design the optimal power allocation to achieve the balance between security, reliability, and covertness. }{When Alice has the knowledge of Eve, \emph{\textbf{AN can be designed as adversarial attack}}. As mentioned in Section \ref{subsection_JSC}, deep learning models  are highly sensitive to adversarial attacks, and the attacker can use the adversarial attack to mislead the semantic receiver of Bob. Here,  we attempt to explore the positive side of adversarial attacks, and the AN generation module generates  adversarial attacks to mislead  the deep learning models of Eve.}



\subsection{Denoising Module}

As shown in Fig. \ref{SC_eavesdropping}, the actively introduced AN and the passively introduced channel noise can be regarded as the forward process of the diffusion model.  This drives us to use the reverse process of the diffusion model to remove the artificial noise and the channel noise from the legitimate link. Thus, we install a denoising module in front of Bob's semantic receiver, which primarily consists of two key steps: time-step estimation and noise estimation.

\subsubsection{Time-step Estimation}
In the reverse process of the diffusion model, the core is to train a noise estimation neural network, where the input consists of noisy signals and the corresponding time-step, as shown in  Fig. \ref{diffusion}. In the training process, the noisy signal at the $t$-th time-step is generated by the  variance
schedule $\beta_t$.  In the testing process, the noisy signal is the received signal of Bob, which is affected by the AN and the channel noise.  To ensure that the reverse process of diffusion models effectively removes noise from the received signal, researchers can estimate the time-step $t$ such that the signal-to-noise ratio (SNR) of the received signal at Bob is as consistent as possible with the SNR obtained by adjusting with the variance schedule $\beta_t$.

\subsubsection{Noise Estimation}
Once the time-step $t$ is estimated, the received signal at Bob and the time-step $t$ can be input into the noise estimation neural network, which will then produce the denoised signal. For the noise estimation neural network, it can be  trained using the loss function that calculates an error measure (e.g., MSE) between the true diffused noise  and the approximated noise. Specifically, the true diffused noise  is generated  by adding Gaussian noise with the variance schedule $\beta_t$. The approximated noise refers to the output of the noise estimation neural network.

\subsection{Case Study for  Eavesdropping Attack Scenario}

The proposed eavesdropping-aware semantic communication framework can be applied various eavesdropping  attack scenario. Here, we conduct a case study to delve into the proposed framework. Specifically, we consider a scenario where Alice and Bob are performing an image reconstruction task, while Eve attempts to launch an attribute inference attack to retrieve the labels of the images being transmitted by Alice.


The part A of Fig. \ref{fashion_eavesdropping} demonstrates the trade-offs between the communication MSE, the privacy leakage mutual information (MI), and the perceptibility MSE in the scenario where Alice has no knowledge of Eve.  Clearly, the AN generation module ensures that the privacy leakage MI of the eavesdropping link approaches zero. Additionally, the denoising module significantly reduces the communication MSE of the Alice-Bob link, thereby ensuring the quality of semantic communication. Furthermore,
the perceptibility MSE is well below the upper bound, indicating that the insertion of the AN generation module effectively ensures covertness. The part B of Fig. \ref{fashion_eavesdropping} shows  the trade-offs between the communication MSE, the classification accuracy of Eve, and the perceptibility MSE in the scenario where Alice has the knowledge of Eve. We can see that the insertion of the AN generation module causes Eve's classification accuracy to drop to around $0.2$, which is nearly an ineffective level.  The insertion of the denoising module enhances the semantic communication quality of the Alice-Bob link, outperforming the original semantic communication system. Meanwhile, we observe that the perceptibility MSE consistently remains below $0.1$,\footnote{Note that we set the output power of the semantic transmitter to $1$.} indicating that it is difficult for the eavesdropper to determine whether Alice has inserted AN generation module.

\begin{figure*}[tbp]
  \centering
  \includegraphics[width=5.6in]{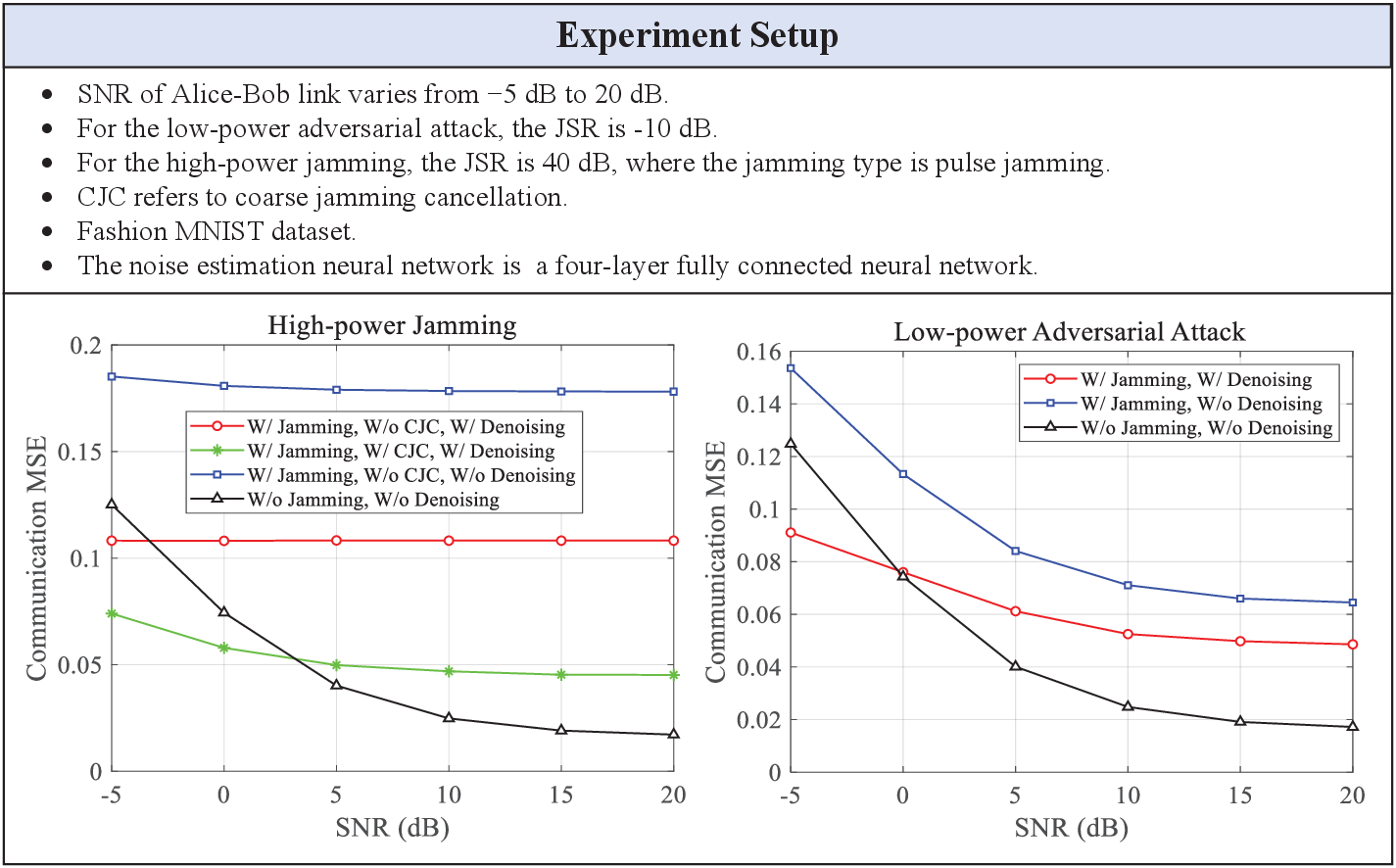}\\
  \caption{  Performance evaluation of the proposed jamming-aware semantic communication framework via diffusion models.}  \label{fashion_jamming}
\end{figure*}

\section{Jamming-aware Semantic Communication Framework Via Diffusion Models} \label{SectionV}

To address the jamming problem of semantic communication, as shown in Fig. \ref{SC_jamming}, we propose a novel jamming-aware semantic communication framework via diffusion models. Specifically, if the jamming signal is a high-power jamming, then the traditional jamming cancellation technique is first used to cancel the jamming coarsely, and then the denoising module is designed to eliminate the residual jamming and the channel noise. On the contrary, if the jamming signal is a low-power adversarial attack, then the denoising module is directly applied to cancel the adversarial attack and the  channel noise.
\subsection{Jamming Identification and Coarse  Jamming Cancellation}

The goal of jamming identification is to determine the type of jamming signal emitted by the jammer. Traditional  jamming identification method such as decision tree has been studied widely, which  can be also used in our framework. For example, we can first distinguish between adversarial attacks and high-power jamming signals by analyzing JSR of the received signal. Then, the expert features such as the frequency characteristics, power spectrum, and higher-order cumulant can be further used to obtain the type of the high-power jamming signal. In addition, deep learning technology demonstrated the powerful capabilities in classification problems. Jamming  identification is essentially a classification problem. Thus, we can identify the jamming type using the trained network.
Once the identified jamming signal is classified as the high-power jamming signal, a coarse jamming cancellation  should be performed on the received signal. Since the type of jamming is already known, the simplest approach is to reconstruct and subtract the jamming. Specifically,
using known parameters of the jamming (such as power spectrum, frequency, phase, or modulation), the jamming signal is first reconstructed. The reconstructed jamming is then subtracted from the received signal to retrieve the original signal.


\subsection{Fine Denoising via Diffusion Models}

Similar to the eavesdropping case, we regard the jamming actively generated by the jammer and the passively introduced channel noise  as the forward process of the diffusion models. Then, the inverse process of the diffusion models is used to remove the jamming (or the residual jamming) and the channel noise. To this end, we first need to determine the time-step of the diffusion models, as it is one of the inputs to the noise estimation neural network. A feasible approach is to find a time-step $t$ such that the SNR of the forward process is as close as possible to the signal-to-interference-plus-noise ratio (SINR) of the input signal of the denoising module. Then, both the estimated time-step and the received signal (or the output of the coarse jamming cancellation) are fed into the trained noise estimation neural network. Finally, we can obtain the denoised signal by subtracting the estimated noise from the input signal of the noise estimation neural network.

\subsection{Case Study for  Jamming Attack Scenario}

The proposed jamming-aware semantic communication framework can be applied various jamming  attack scenario.
 Here,  we delve into the jamming-aware semantic communication framework by carrying out the case study. Specifically, we consider a scenario where Alice and Bob are performing an image reconstruction task, while Eve transmits jamming signals  to disrupt Bob's reconstruction process.

The left part of Fig. \ref{fashion_jamming} gives the communication MSE in the case of high-power jamming, where the JSR is $40$ dB and the jamming type is the pulse jamming. First, we can observe that the high-power jamming causes Bob to fail in performing the image reconstruction task. Then, it can be observed that directly applying the reverse process of the diffusion model can reduce the communication MSE, but it still remains at a relatively high level. Furthermore, we find that by first performing coarse jamming cancellation and then using the diffusion model for the fine denoising, the communication MSE can be reduced to a level close to that of no jamming. The right part of Fig. \ref{fashion_jamming} shows the communication MSE in the case of low-power adversarial attack, where the JSR is $-10$ dB. We can observe that due to the sensitivity of the deep neural networks, the adversarial attack sent by the jammer causes a sharp deterioration in the communication MSE of the Alice-Bob link. By using the reverse process of the diffusion models, the communication MSE is significantly improved, with the enhancement being more pronounced as the SNR decreases. This result demonstrates that the diffusion models is highly effective in mitigating the adversarial attacks generated by the jammer and the channel noise.

\section{Challenges and Future Directions} \label{SectionVI}

In this section, we present the challenges and future
directions of the diffusion-enabled semantic transmission.

\subsection{Security Design in Complex Channel Conditions}

Wireless environments are often complex, consisting of additive Gaussian white noise channels and various types of fading channels. Our work demonstrates how diffusion models can be used to prevent eavesdropping or jamming when the channel is additive noise. In fading channel scenarios, the received signal is affected by both the multiplicative fading channel and additive noise. Intuitively, due to the impact of the multiplicative channel, the signal transmission process does not fully align with the forward process of the diffusion model. Although existing literature has matched the signal transmission process with the forward process of the diffusion model by channel equalization\cite{wu2024cddm}, how to extend the existing approach to the proposed  diffusion-enabled secure semantic transmission still requires further research.

\subsection{Security Design for Training Phase}

The proposed diffusion-based framework is proposed for the deployment phase of the semantic communication. However, the study of the secure semantic communication  for the training phase is still in its infancy. In semantic communication systems, attacks may also occur during the model training process such as backdoor attacks and poisoning attacks. Poisoning attacks typically aim to cause a model to produce incorrect or inappropriate outputs by injecting poisoned or noisy data into its training process. In contrast, backdoor attacks are designed to trigger the model to produce specific, predetermined outputs when exposed to certain inputs. The inability and susceptibility to such attacks can undermine the system¡¯s ability and make the system fail to achieve the certain goals in goal-oriented semantic communication systems. Our work demonstrates the strong potential of diffusion models in removing channel noise and jamming. Since attacks such as backdoor attacks and poisoning attacks can be viewed as a special form of jamming, exploring the use of diffusion models to achieve secure semantic communication in the training phase is an interesting direction for future research.

\subsection{Cross-layer Security Design}

Cross-layer security design for semantic communication focuses on integrating security mechanisms across multiple layers of the system. Our work proposes a novel diffusion-based secure semantic communication from the perspective of the physical layer. In semantic communication, each layer plays a unique role in the processing of data. For instance, security measures at one layer might affect the performance of higher layers. The interaction between layers must be carefully considered to ensure that security measures are effective without compromising efficiency, particularly in real-time systems.

\subsection{Low-complexity Security Design}

One of the drawbacks of the diffusion models is that they require a substantial amount of time for training. This poses a significant challenge for resource-constrained edge devices in wireless networks. To reduce complexity of the security scheme via diffusion models, future works can  leverage techniques such as model pruning or distributed learning to address the resource limitations of edge devices. In addition, task-agnostic diffusion model design is also  a promising direction, as this means we do not need to customize a specific diffusion model for each task to ensure security.

\section{Conclusion} \label{SectionVII}

This paper exploited  how diffusion models can be used to address eavesdropping and jamming issues in semantic communication systems.
Specifically, we proposed converting both the artificially introduced noise and the natural channel noise into the forward process of  diffusion models, and then utilizing the reverse process of  diffusion models to remove the noises and generate the meaningful  information. Through case studies in the semantic eavesdropping and jamming scenarios,  it verified that the proposed diffusion-based framework can ensure the security in semantic communication systems. The potential future works for the diffusion-enabled framework are expected to encompass the security design for the complex channel conditions, the training phase, the cross-layer, and the low-complexity.

\bibliographystyle{IEEEtran}
\bibliography{ref}

\end{document}